\documentstyle[12pt,epsf]{article}
\begin{document}
\newcommand{\be}{\begin{equation}}
\newcommand{\ee}{\end{equation}}
\newcommand{\ba}{\begin{array}}
\newcommand{\ea}{\end{array}}
\newcommand{\bea}{\begin{eqnarray}}
\newcommand{\eea}{\end{eqnarray}}

\title{Superintegrability on the two dimensional hyperboloid II}
\author{{\bf E. G. Kalnins} \\
{\sl Department of Mathematics and Statistics,}\\
{\sl University of Waikato,}\\
{\sl Hamilton, New Zealand.}\\
{\bf W.\ Miller, Jr.}\\
{\sl School of Mathematics, University of Minnesota,}\\
{\sl Minneapolis, Minnesota,
55455, U.S.A.}\\
{\bf Ye. M. Hakobyan and G.\ S.\ Pogosyan}\\
{\sl Laboratory of Theoretical Physics,}\\
{\sl Joint Institute for Nuclear Research,}\\
{\sl Dubna, Moscow Region 141980, Russia}}
\date{\today}
\maketitle

\vspace{0.3cm}

{\large {\bf Abstract}}

\vspace{0.3cm}

\noindent
This work is devoted to the investigation of the quantum
mechanical systems on the two dimensional hyperboloid which admit
separation of variables in at least two coordinate systems.
Here we consider two potentials introduced in a paper of
C.P.Boyer, E.G.Kalnins and P.Winternitz, which
haven't yet been studied. We give an example of an interbasis
expansion and work out the structure of the quadratic algebra
generated by the integrals of motion.

\section{Introduction}

Superintegrable systems on the two-dimensional hyperboloid were
introduced and developed in the papers \cite{BKW,GPSI,KMJP1}. In
distinction to the cases of two-dimensional Euclidean space and the
two-sphere, the classification of superintegrable systems on the
hyperboloid is difficult. To date only the four potentials studied in
\cite{KMJP1} and two more listed in \cite{BKW} are known.
In the present paper two potentials are considered,
which were constructed in the work \cite{BKW} but have not previously
been investigated.  These potentials both have only a finite number of
bound states. At this point we have treated all the potentials that
arise by restriction from hermitean hyperbolic space. We follow the approach of \cite{KMJP1}, which
contains an introduction and motivation.

The two dimensional hyperboloid is characterized via the cartesian
coordinates $\omega_0,\omega_1,\omega_2$ where $\omega ^2_0-\omega
^2_1-\omega^2_2=1, \omega _0>1.$ The requirement $\omega_0 >1$ means that we
consider only upper sheet of the double-sheet hyperboloid. Throughout this
paper we will consider the Schr\"odinger equation on the hyperboloid in the
form ($\hbar=m=1$)
\begin{equation}
\label{SCH}
H\Psi \equiv \left(-\frac{1}{2}\Delta_{LB} + V\right)\Psi = E \Psi
\end{equation}
where $V$ is a potential function and the Laplace-Beltrami operator
$\Delta_{LB}$ is written as
%%%%%%%%%%%%%%%%%%%%%%%%%%%%%%%%%%%%%%%%%%%%%%%%%
\begin{equation}
\Delta_{LB} = K^2_3 + K^2_2 - M^2_1.
\end{equation}
Here $K_3, K_2, M_1$ generate the Lie algebra $so(2,1)$
\cite{WLS,KM}
%%%%%%%%%%%%%%%%%%%%%%%%%%%%%%%%%%%%%%%%%%%%%%%%%
\begin{eqnarray}
K_3 = \omega _0\partial _{\omega _1}+\omega _1\partial _{\omega _0},\quad
K_2 = \omega _0\partial _{\omega _2}+\omega _2\partial _{\omega _0}, \quad
M_1 = \omega_1\partial_{\omega_2} - \omega_2 \partial_{\omega _1}.
\end{eqnarray}
and
\begin{eqnarray}
[K_3,K_2]=M_1, \quad
[K_2,M_1]=-K_3, \quad
[K_3,M_1]=K_2  \quad
\end{eqnarray}
%%%%%%%%%%%%%%%%%%%%%%%%%%%%%%%%%%%%%%%%%%%%%%%%%
The Schr\"odinger equation (\ref{SCH}) for $V=0$
separates in nine coordinate systems \cite{OLE}. Introduction of a
potential breaks the symmetry and, in general, reduces the number of
coordinate systems permitting separability, usually to zero.
We consider the following two potentials (see Table), constructed in  \cite{BKW}, for which
(\ref{SCH}) is superintegrable.

\hspace{1.5cm}

%%%%%%%%%%%%%%%%%%%%%begin of the table%%%%%%%%%%%%%%%%%%%%%%
%\newpage

$Table$

\hspace{1.0cm}

\begin{tabular}{|l|l|}
\hline
Potential $V(\omega)$ & Coordinate system \\ \hline
& Equidistant \\
&   Elliptic-parabolic  \\
$V_1 = \frac{\alpha^2}{\omega_2^2}
-\frac{\gamma^2}{(\omega_0-\omega_1)^2}
+\beta^2\frac{\omega_0 + \omega_1} {(\omega_0-\omega_1)^3}$ &  \\
& Hyperbolic-parabolic\\
& Horicyclic \\ \hline
& Equidistant \\
$V_2 = \frac{\alpha^2}{\omega_2^2}
+\gamma^2\frac{\omega_0\omega_1}{(\omega_0^2+\omega_1^2)^2}$ & \\
& Semi-Hyperbolic \\
$+(\alpha^2-\beta^2)\frac{\omega_0^2-\omega_1^2}{(\omega_0^2+\omega_1^2)^2}
$ & \\
&  \\ \hline
\end{tabular}
%%%%%%%%%%%%%%%%%%%%%end of the table%%%%%%%%%%%%%%%%%%%%%%
\\

\vspace{1cm}

\hspace{0.5cm}
Recall that (\ref{SCH}) is {\it superintegrable} for a given potential
$V$ if it is separable simultaneously in at least two coordinate systems.
%\newpage

\section {First Potential}

The first considered potential is
%%%%%%%%%%%%%%%%%%%%%%%%%%%%%%%%%%%%%%%%%%
\begin{equation}
\label{F1}
V_1 = \frac{\alpha^2}{\omega_2^2}
-\frac{\gamma^2}{(\omega_0-\omega_1)^2}
+\beta^2\frac{\omega_0 + \omega_1} {(\omega_0-\omega_1)^3}
\end{equation}
%%%%%%%%%%%%%%%%%%%%%%%%%%%%%%%%%%%%%%%%%%
where $\alpha, \beta, \gamma$ are positive constants .
The corresponding Schr\"odinger equation admits separable solutions in four
coordinate systems: equidistant, elliptic--parabolic, hyperbolic--parabolic
and horicyclic.

\subsection{Solutions of the Schr\"odinger equation}

\vspace{0.5cm} {\it\underline{1.1 Equidistant coordinates.}}
In this coordinate system
\[
\omega_0 = \cosh \tau_1 \cosh \tau_2, \,\,\,\, \omega_1 = \cosh \tau_1 \sinh \tau_2, \,\,\,\,
\omega_2 = \sinh \tau_1
\]
$[\tau_1,\tau_2 \in (-\infty, \infty)]$ the potential $V_1$ has the form
%%%%%%%%%%%%%%%%%%%%%%%%%%%%%%%%%%
\begin{equation}
\label{FE1}
V_1(\tau_1,\tau_2) =
\frac{\alpha^2}{\sinh^2 \tau_1}
+\frac{1}{\cosh^2 \tau_1}
\frac{\beta^2-\gamma^2(\cosh \tau_2-\sinh \tau_2)^2}
{(\cosh \tau_2-\sinh \tau_2)^4}
\end{equation}
%%%%%%%%%%%%%%%%%%%%%%%%%%%%%%%%%%
After putting
%%%%%%%%%%%%%%%%%%%%%%%%%%%%%%%%%%
\begin{equation}
\label{FE2}
\Psi(\tau_1,\tau_2) = (\cosh \tau_1)^{-1/2} S_1(\tau_1) S_2(\tau_2)
\end{equation}
%%%%%%%%%%%%%%%%%%%%%%%%%%%%%%%%%%
we come to the system of equations:
%%%%%%%%%%%%%%%%%%%%%%%%%%%%%%%%%%
\begin{eqnarray}
\label{FE3}
\frac{d^2 S_2}{d \tau_2^2} &+& \left[ -\mu^2-2
\beta^2 e^{4 \tau_2}+2\gamma^2 e^{2 \tau_2}\right] S_2=0
\\[2mm]
\label{FE4}
\frac{d^2 S_1}{d \tau_1^2} &+& \left[(2 E-\frac{1}{4}) +
\frac{\mu^2-\frac{1}{4}}{\cosh^2\tau_1}
- \frac{2\alpha^2}{\sinh^2 \tau_1} \right] S_1 = 0
\end{eqnarray}
%%%%%%%%%%%%%%%%%%%%%%%%%%%%%%%%%%
where $\mu$ is the equidistant separation constant.
The first equation (\ref{FE3}) could be considered as
 a one dimensional Shr\"odinger equation for the Morse potential \cite{LN}
and the orthonormalised solution is given by the expression:
%%%%%%%%%%%%%%%%%%%%%%%%%%%%%%%%%%
\begin{eqnarray}
S_2(\tau_2) \equiv S^{(\beta, \mu)}_{m} (z)
&=& \sqrt{\frac{2\mu \Gamma(m+\mu+1)}{m!\Gamma^2(\mu+1)}}
e^{-z/2}z^{\mu/2}\,
_2F_1( -m, \mu+1; z)
\nonumber
\\[2mm]
\label{FE5}
&=& \sqrt{\frac{2\mu m!}{\Gamma(m+\mu+1)}} e^{-z/2} z^{\mu/2}
L_m^{\mu}(z),\quad z=\sqrt{2}\beta e^{2 \tau_2}
\eea
%%%%%%%%%%%%%%%%%%%%%%%%%%%%%%%%%%
where $L_m^{\mu}(z)$ are the Laguerre polynomials \cite{EMOT}.
The separation constant is quantized as
%%%%%%%%%%%%%%%%%%%%%%%%%%%%%%%%%%
\bea
\label{FE6}
\mu = -2 m-1+\frac{\gamma^2}{\sqrt{2}\beta},\quad 0\leq m\leq
\left[\frac{1}{2}\left(\frac{\gamma^2}{\sqrt{2}\beta}-1\right)\right]
\eea
%%%%%%%%%%%%%%%%%%%%%%%%%%%%%%%%%%
The second equation (\ref{FE4}) represents
the modified P\"oschl--Teller equation~\cite{KMJP1,FWO}.
The orthonormalised wave function is  given by:
%%%%%%%%%%%%%%%%%%%%%%%%%%%%%%%%%%
\begin{eqnarray}
S_1(\tau_1) &\equiv& S^{(\alpha, \mu)}_{n} (\tau_1)
= \sqrt{\frac{2 (\mu-\sqrt{2\alpha^2+1/4}-2n-1)\Gamma(\mu-n)n!}
{\Gamma(\mu-\sqrt{2\alpha^2+1/4}-n)\Gamma(1+n+\sqrt{2\alpha^2+1/4})}}
\nonumber
\\[2mm]
\label{FE7}
&\times&
(\sinh \tau_1)^{\frac{1}{2}+\sqrt{2\alpha^2+1/4}}
(\cosh \tau_1)^{\frac{1}{2}-\mu}
P^{(\sqrt{2\alpha^2+1/4}, -\mu)}_n(\cosh 2 \tau_1),
\end{eqnarray}
%%%%%%%%%%%%%%%%%%%%%%%%%%%%%%%%%%
with $n=0,1,\dots \left[\frac12\left(\mu-1-\sqrt{2\alpha^2
+ \frac14}\right)\right]$, where $P^{(\alpha, \beta)}_n(x)$
is the Jacobi polynomial \cite{EMOT}. The quantized energy is
%%%%%%%%%%%%%%%%%%%%%%%%%%%%%%%%%%
\bea
E_N &=& -\frac{1}{2}(\mu-\sqrt{2\alpha^2+1/4}-2n-1)^2+\frac{1}{8}
\nonumber
\\[2mm]
\label{FE8}
&=&-\frac{1}{2}\left(2N+2+\sqrt{2\alpha^2+1/4}
-\frac{\gamma^2}{\sqrt{2}\beta}\right)^2
+\frac{1}{8}
\eea
%%%%%%%%%%%%%%%%%%%%%%%%%%%%%%%%%%
where $N=m+n$ is the principal quantum number and the bound states
occur for
%%%%%%%%%%%%%%%%%%%%%%%%%%%%%%%%%%
\be
\label{N}
0\leq N \leq \left[ \frac{1}{2}\left(\frac{\gamma^2}{\sqrt{2}\beta}
-\sqrt{2\alpha^2+1/4}-2\right)\right]
\ee
%%%%%%%%%%%%%%%%%%%%%%%%%%%%%%%%%%
The orthonormalized total wave function $\Psi_{n m}(\tau_1, \tau_2)$
is given by (\ref{FE2}), (\ref{FE5}) and (\ref{FE7}).

The symmetry operator describing this coordinate system is
%%%%%%%%%%%%%%%%%%%%%%%%%%%%%%%%%%
\begin{eqnarray}
 L_1 \Psi_{n m}(\tau_1, \tau_2) &\equiv&
 \left[K_3^2
-2\beta^2 \left(\frac{\omega_0+\omega_1}{\omega_0-\omega_1}\right)^2
+2\gamma^2 \frac{\omega_0+\omega_1}{\omega_0-\omega_1}\right]\Psi_{n m}(\tau_1, \tau_2)
\nonumber
\\[2mm]
\label{FE10}
&=&\left(-2 m-1+\frac{\gamma^2}{\sqrt{2}\beta}\right)^2
\Psi_{n m}(\tau_1, \tau_2)
\end{eqnarray}
%%%%%%%%%%%%%%%%%%%%%%%%%%%%%%%%%%

\vspace{0.5cm}
\underline{{\it 1.2 Horicyclic coordinates}}.
In the horicyclic coordinates
%%%%%%%%%%%%%%%%%%%%%%%%%%%%%%%%%%%%%%%%%%
\bea
\label{FHC1}
\omega_0 = \frac{x^2+y^2+1}{2y},
\,\,\,\,
\omega_1 = \frac{x^2+y^2-1}{2y},
\,\,\,\,
\omega_2 = \frac{x}{y},
\eea
%%%%%%%%%%%%%%%%%%%%%%%%%%%%%%%%%%%%%%%%%%
$[y>0, x\in (-\infty, \infty)]$ the potential $V_1$ is
%%%%%%%%%%%%%%%%%%%%%%%%%%%%%%%%%%
\begin{eqnarray}
\label{FHC2}
V_1 (x, y) = y^2\left[
\frac{\alpha^2}{x^2}+\beta^2(x^2+y^2)-\gamma^2\right]
\end{eqnarray}
%%%%%%%%%%%%%%%%%%%%%%%%%%%%%%%%%%
and the Schr\"odinger equation has the following form
%%%%%%%%%%%%%%%%%%%%%%%%%%%%%%%%%%
\begin{eqnarray}
\label{FHC3}
-\frac{1}{2}y^2\bigg[
\frac{\partial^2}{\partial x^2}-\frac{2 \alpha^2}{x^2}-2\beta^2x^2+
\frac{\partial^2}{\partial y^2}-2\beta^2y^2+ 2\gamma^2\bigg]\Psi(x,y)
=E\Psi(x,y)
\end{eqnarray}
%%%%%%%%%%%%%%%%%%%%%%%%%%%%%%%%%%
Via putting
%%%%%%%%%%%%%%%%%%%%%%%%%%%%%%%%%%
\begin{eqnarray}
\label{FHC}
\Psi (x,y) =\psi_1(x)\psi_2(y)
\end{eqnarray}
%%%%%%%%%%%%%%%%%%%%%%%%%%%%%%%%%%
it admits a separation
%%%%%%%%%%%%%%%%%%%%%%%%%%%%%%%%%%
\begin{eqnarray}
\label{FHC4}
\frac{d^2 \psi_1}{d x^2}+2\left[
\gamma^2(\lambda_1+1)-\beta^2x^2-\frac{ \alpha^2}{x^2}\right]\psi_1=0
\\[2mm]
\label{FHC5}
\frac{d^2 \psi_2}{d y^2}+2\left[
\gamma^2(\lambda_2-1)-\beta^2 y^2+\frac{E}{y^2}\right]\psi_2=0
\end{eqnarray}
%%%%%%%%%%%%%%%%%%%%%%%%%%%%%%%%%%
where $\lambda_1$ and $\lambda_2$ are the horicyclic separation constants
with the relation: $\lambda_1+\lambda_2=1$.

The orthonormalized solutions of the equations (\ref{FHC4}), (\ref{FHC5})
for $(- 2E+1/4) > 0$ are
%%%%%%%%%%%%%%%%%%%%%%%%%%%%%%%%%%
\begin{eqnarray}
\psi_1(x)&\equiv&\psi_{n_1}^{(\alpha,\beta)}(x)=
\sqrt{\frac{n_1!(\sqrt{2}\beta)^{1/2}}{\Gamma(n_1+\sqrt{2\alpha^2+1/4}+1)}}
\nonumber
\\[2mm]
\label{FHC6}
&\times&e^{-\frac{\beta x^2}{\sqrt{2}}}
(\sqrt{\sqrt{2}\beta x^2})^{\frac{1}{2}+\sqrt{2\alpha^2+1/4}}
L_{n_1}^{\sqrt{2\alpha^2+1/4}}(\sqrt{2}\beta x^2)
\end{eqnarray}
%%%%%%%%%%%%%%%%%%%%%%%%%%%%%%%%%%
\begin{eqnarray}
\psi_2(y)&\equiv&\psi_{n_2}^{(\gamma,\beta)}(y)=
\sqrt{\frac{n_2!(\sqrt{2}\beta)^{1/2}}{\Gamma(n_2+\sqrt{-2E+1/4}+1)}}
\nonumber
\\[2mm]
\label{FHC8}
&\times&e^{-\frac{\beta y^2}{\sqrt{2}}}
(\sqrt{\sqrt{2}\beta y^2})^{\frac{1}{2}+\sqrt{-2E+1/4}}
L_{n_2}^{\sqrt{-2E+1/4}}(\sqrt{2}\beta y^2).
\end{eqnarray}
%%%%%%%%%%%%%%%%%%%%%%%%%%%%%%%%%%
The separation constants $\lambda_1$, $\lambda_2$ are quantized as:
%%%%%%%%%%%%%%%%%%%%%%%%%%%%%%%%%%
\begin{eqnarray}
\label{FHC7}
\lambda_1=\frac{\sqrt{2}\beta}{\gamma^2}(2 n_1+\sqrt{2 \alpha^2+1/4}+1)-1;\quad
\lambda_2=\frac{\sqrt{2}\beta}{\gamma^2}(2 n_2+\sqrt{-2E+1/4}+1)+1
\end{eqnarray}
%%%%%%%%%%%%%%%%%%%%%%%%%%%%%%%%%%
and according to the relation $\lambda_1+\lambda_2=1$, we come to
the energy spectrum as in (13). The operator characterizing the
separation in horicyclic coordinates is:
%%%%%%%%%%%%%%%%%%%%%%%%%%%%%%%%%%
\begin{eqnarray}
 L_2 \Psi_{n_1 n_2} (x,y) &\equiv&
\left[(K_2-M_1)^2-\frac{2\beta^2 \omega_2^2}{(\omega_0-\omega_1)^2}
-\frac{2\alpha^2 (\omega_0-\omega_1)^2}{\omega_2^2} +2 \gamma^2
\right]\Psi_{n_1 n_2} (x,y)
\nonumber
\\[2mm]
\label{FHC11}
&=&-\left[2
\sqrt{2}\beta(2 n_1+\sqrt{2 \alpha^2+1/4}+1)+2 \gamma^2\right]
\Psi_{n_1 n_2} (x,y)
\end{eqnarray}
%%%%%%%%%%%%%%%%%%%%%%%%%%%%%%%%%%

\vspace{0.5cm}

{\it \underline{1.3 Elliptic--parabolic coordinates}}.
In this coordinate system
%%%%%%%%%%%%%%%%%%%%%%%%%%%%%%%%%%
\begin{eqnarray}
\label{FEP0}
\omega_0 = {\cosh^2a + \cos ^2\theta \over 2\cosh a\cos\theta},
\,\,\,\,
\omega_1 = {\sinh^2a - \sin^2\theta \over 2\cosh a\cos\theta},
\,\,\,\,
\omega_2 = \tanh a \tan\theta,
\end{eqnarray}
%%%%%%%%%%%%%%%%%%%%%%%%%%%%%%%%%%
$[a > 0, \, \theta\in(-\frac{\pi}{2},
\frac{\pi}{2})]$
the potential $V_1$ has the form:
%%%%%%%%%%%%%%%%%%%%%%%%%%%%%%%%%%
\begin{eqnarray}
\label{FEP1}
V_1 (a, \theta) &=&
\frac{\cosh^2a\cos^2\theta}{\cosh^2a-\cos^2\theta}
\bigg[\beta^2(\cosh^2a \sinh^2 a+\cos^2\theta\sin^2\theta)
\nonumber
\\[2mm]
&-&\gamma^2(\cosh^2 a-\cos^2\theta)+
\left.\alpha^2\left(\frac{1}{\sinh^2a}+\frac{1}{\sin^2\theta}\right)\right]
\end{eqnarray}
%%%%%%%%%%%%%%%%%%%%%%%%%%%%%%%%%%
The Shr\"odinger equation is:
%%%%%%%%%%%%%%%%%%%%%%%%%%%%%%%%%%
\begin{eqnarray}
\label{FEP2}
&-&\frac{1}{2}\frac{\cosh^2a\cos^2\theta}{\cosh^2a-\cos^2\theta}
\bigg[\frac{\partial^2}{\partial a^2}-2 \beta^2\cosh^2a\sinh^2a
+ 2\gamma^2\cosh^2 a-\frac{2\alpha^2}{\sinh^2a}\nonumber
\\[2mm]
&+&\frac{\partial^2}{\partial \theta^2}
-2\beta^2\cos^2\theta\sin^2\theta- 2\gamma^2\cos^2\theta
-\frac{2\alpha^2}{\sin^2\theta}
\bigg]\Psi(a,\theta) =E\Psi(a,\theta)
\end{eqnarray}
%%%%%%%%%%%%%%%%%%%%%%%%%%%%%%%%%%
Putting for the wave function $\Psi(a, \theta) = S(a) S(\theta)$,
after separation of variables we get two identical equations:
%%%%%%%%%%%%%%%%%%%%%%%%%%%%%%%%%%
\bea
\label{FEP3}
\frac{d^2 S(\rho)}{d\rho^2}
+ \bigg[\lambda-2\beta^2\cosh^2\rho\sinh^2\rho
+ 2 \gamma^2\cosh^2\rho
-\frac{2\alpha^2}{\sinh^2\rho}
-\frac{2E}{\cosh^2 \rho}\bigg]S(\rho)=0
\end{eqnarray}
%%%%%%%%%%%%%%%%%%%%%%%%%%%%%%%%%%
where $\lambda$ is the elliptic--parabolic separation constant
and $\rho\equiv a, i\theta$.
After changing the variables  $x=\cosh^2\rho$ in eq. (\ref{FEP3}),
we obtain
%%%%%%%%%%%%%%%%%%%%%%%%%%%%%%%%%%
\bea
\label{FEP4}
4x(x-1)\frac{d^2 S}{d x^2}+2(2x-1)\frac{d S}{d x}
+\left[
\lambda-2\beta^2x(x-1)+2\gamma^2x-\frac{2\alpha^2}{x-1}-\frac{2 E}{x}
\right]S=0
\end{eqnarray}
%%%%%%%%%%%%%%%%%%%%%%%%%%%%%%%%%%
Thus the region $x\in [1,\infty]$ in eq. (\ref{FEP4}) belongs to the wave
function $S(a)$ and $x\in [0,1]$ to the wave function $S(\theta)$.
Putting
%%%%%%%%%%%%%%%%%%%%%%%%%%%%%%%%%%
\bea
\label{FEP5'}
S(x) = (x-1)^{s}x^{t}e^{-\beta x /\sqrt{2}}G(x),
\end{eqnarray}
%%%%%%%%%%%%%%%%%%%%%%%%%%%%%%%%%%
where
%%%%%%%%%%%%%%%%%%%%%%%%%%%%%%%%%%
\bea
\label{FEP5}
s=\frac{1}{4}+\frac{1}{\sqrt{2}}\sqrt{\alpha^2+\frac{1}{8}},
\quad t=\frac{1}{4}+\frac{1}{\sqrt{2}}\sqrt{-E+\frac{1}{8}}
\end{eqnarray}
%%%%%%%%%%%%%%%%%%%%%%%%%%%%%%%%%%
we get
%%%%%%%%%%%%%%%%%%%%%%%%%%%%%%%%%%
\bea
\label{FEP7}
\frac{d^2 G}{d x^2} &+& \frac{1}{2}\bigg[ \frac{1+4 t}{x}
+\frac{1+4 s}{x-1}- \frac{4 \beta}{\sqrt{2}} \bigg]\frac{d G}{d x}
\\[2mm]
&+&\frac{1}{4}\bigg\{\frac{[2\gamma^2-4\beta(1+2(t+s))/\sqrt{2}]x
+\nu+\sqrt{2}\beta(1+4 t)+4(t+s)^2}{x(x-1)}\bigg\}G=0.
\nonumber
\end{eqnarray}
%%%%%%%%%%%%%%%%%%%%%%%%%%%%%%%%%%
If we now substitute
%%%%%%%%%%%%%%%%%%%%%%%%%%%%%%%%%%
\bea
\label{FEP8}
G(x)=\prod\limits_{i=1}^{N}(x-\theta_i)
\end{eqnarray}
%%%%%%%%%%%%%%%%%%%%%%%%%%%%%%%%%%
and take into account (\ref{FEP5}),
we find that $\theta_i$ satisfies the equation:
%%%%%%%%%%%%%%%%%%%%%%%%%%%%%%%%%%
\bea
2\theta_i(1-\theta_i)\left(\sum^{N}_{\scriptstyle k=1 \atop\scriptstyle k\not=i}
\frac{1}{\theta_k-\theta_i}
+\frac{\beta}{\sqrt{2}}\right)
+2(1-\theta_i)N+
\frac{\sqrt{2}\gamma^2}{4\beta}\theta_i
\nonumber
\\[2mm]
\label{FEP11}
+\frac{\gamma^2}{\sqrt{2}\beta}+
\sqrt{2\alpha^2+\frac{1}{4}}+1&=&0.
\end{eqnarray}
%%%%%%%%%%%%%%%%%%%%%%%%%%%%%%%%%%
The quantization for the energy is given via:
%%%%%%%%%%%%%%%%%%%%%%%%%%%%%%%%%%
\bea
\label{FEP9}
\sqrt{-2 E+\frac{1}{4}}+\sqrt{2\alpha^2+\frac{1}{4}}
+2N + 2
- \frac{\gamma^2}{\sqrt{2}\beta}=0
\end{eqnarray}
%%%%%%%%%%%%%%%%%%%%%%%%%%%%%%%%%%
and we obtain the expression (\ref{FE8}).
The separation constant $\lambda$ is:
%%%%%%%%%%%%%%%%%%%%%%%%%%%%%%%%%%
\bea
\label{FEP10}
\lambda =
\frac{8\beta}{ \sqrt{2}} \sum_{i=1}^{N}\theta_i
-\left(\frac{\gamma^2}{\sqrt{2}\beta}-1\right)^2+\frac{4\beta}{\sqrt{2}}
\left(1+\sqrt{2 \alpha^2+\frac{1}{4}}\right)-2\gamma^2.
\end{eqnarray}
%%%%%%%%%%%%%%%%%%%%%%%%%%%%%%%%%%
Thus the total solution $\Psi( a, \theta)$ is represented as:
%%%%%%%%%%%%%%%%%%%%%%%%%%%%%%%%%%
\bea
\Psi_{Npq} (a,\theta)&=&
S_{Np}(a)S_{Nq}(\theta)=
(\sinh a\sin\theta)^{\frac12+\sqrt{2\alpha^2+\frac14}}
(\cosh a\cos\theta)^{\frac{\gamma^2}{\sqrt{2}\beta}-\sqrt{2\alpha^2+\frac14}-2N-\frac32}
\nonumber
\\[2mm]
\label{FEP12}
&\cdot&
\exp\left\{-\frac{\beta}{\sqrt{2}} (\cosh^2a+cos^2\theta)\right\}
\prod_{i=1}^{N}(\cosh^2 a-\theta_i)
(\cos^2\theta-\theta_i)
\end{eqnarray}
%%%%%%%%%%%%%%%%%%%%%%%%%%%%%%%%%%
where $p$ and $q$ is the number of zeroes
for the wave functions $S(a)$ and $S(\theta)$
in the regions $[0,1]$,
$[1,\infty]$ correspondingly;
and the total number of zeroes is $N=p+q$.

Eliminating the energy $E$ from equation (\ref{FEP4}), we see that
the additional integral of motion here is
%%%%%%%%%%%%%%%%%%%%%%%%%%%%%%%%%%
\begin{eqnarray}
L_{3} \Psi_{Npq}(a,\theta)
&=&
\frac{1}{\cos^2\theta-\cosh^2 a}\bigg\{
\cosh^2 a
\frac{\partial^2}{\partial a^2}
+\cos^2\theta
\frac{\partial^2}{\partial \theta^2}
-2\beta^2(\cosh^4 a\sinh^2 a
\nonumber
\\[2mm]
&+&
\cos^4 \theta\sin^2 \theta) +2\gamma^2(\cosh^4 a-\cos^4\theta )
-2\alpha^2(\coth ^2 a-\cot^2 \theta)
\bigg\}\Psi_{Npq}(a, \theta)
\nonumber
\\[2mm]
&=&\bigg\{-(K_2-M_1)^2-K_3^2+2\beta^2
\frac{(w_0+w_1)^2+w_2^2}
{(w_0-w_1)^2}+2\alpha^2\left(\frac{w_0-w_1}{w_2}\right)^2
\nonumber
\\[2mm]
\label{FEPl}
&-&4\gamma^2\frac{w_0}{w_0-w_1}
\bigg\}\Psi_{Npq}(a, \theta)
= \lambda \Psi_{Npq}(a, \theta).
\end{eqnarray}
%%%%%%%%%%%%%%%%%%%%%%%%%%%%%%%%%%

\vspace{0.5cm}
\underline{{\it 1.4 Hyperbolic--parabolic coordinates}}.
In this coordinate system
\bea
\label{FH1}
\omega_0 = {\cosh^2b + \cos ^2\theta \over 2\sinh b\sin\theta},
\,\,\,\,
\omega_1 = {\sinh^2b - \sin^2\theta \over 2\sinh b\sin\theta},
\,\,\,\,
\omega_2 = \coth b \cot\theta,
\eea
$[b > 0, \, \theta\in(-\frac{\pi}{2},
\frac{\pi}{2})]$
the potential $V_1$ has the form:
%%%%%%%%%%%%%%%%%%%%%%%%%%%%%%%%%%
\begin{eqnarray}
\label{FH2}
V_1 (b, \theta) &=&
\frac{\sinh^2b\sin^2\theta}{\sinh^2 b+\sin^2\theta}
\left[\beta^2(\sinh^2 b\cosh^2 b+\sin^2\theta\cos^2\theta)\right.
\nonumber
\\[2mm]
&-&\gamma^2 (\sinh^2b+\sin^2\theta)+
\left.\alpha^2\left(\frac{1}{\cos^2\theta}-\frac{1}{\cosh^2 b}\right).
\right]
\end{eqnarray}
%%%%%%%%%%%%%%%%%%%%%%%%%%%%%%%%%%
The Shr\"odinger equation is
%%%%%%%%%%%%%%%%%%%%%%%%%%%%%%%%%%
\begin{eqnarray}
\label{FH3}
&-&\frac{1}{2}\frac{\sinh^2 b\sin^2\theta}{\sinh^2 b+\sin^2\theta}
\bigg[\frac{\partial^2}{\partial b^2}-2 \beta^2\sinh^2 b\cosh^2 b
+2 \gamma^2\sinh^2b +\frac{2\alpha^2}{\cosh^2 b}\nonumber
\\[2mm]
&+&\frac{\partial^2}{\partial \theta^2}
-2\beta^2\sin^2\theta\cos^2\theta
+2 \gamma^2\sin^2\theta-\frac{2\alpha^2}{\cos^2\theta}
\bigg]\Psi(b, \theta)  = E\Psi(b, \theta).
\end{eqnarray}
%%%%%%%%%%%%%%%%%%%%%%%%%%%%%%%%%%
Putting for the wave function $\Psi(b, \theta) = S(b) S(\theta)$,
after separation of variables we get two identical equations:
%%%%%%%%%%%%%%%%%%%%%%%%%%%%%%%%%%
\bea
\label{FH4}
\frac{d^2 S(\rho)}{d\rho^2}
+2\bigg[\frac{\tau}{2}-\beta^2\sinh^2\rho\cosh^2\rho+ \gamma^2\sinh^2\rho
+\frac{\alpha^2}{\cosh^2\rho}
+\frac{E}{\sinh^2 \rho}\bigg] S(\rho)=0
\end{eqnarray}
%%%%%%%%%%%%%%%%%%%%%%%%%%%%%%%%%%
where $\tau$ is the hyperbolic--parabolic separation constant
and $\rho\equiv b, i\theta$.
After changing the variables  $x=\sinh^2\rho$ in eq. (\ref{FH4}),
we come to the equation
%%%%%%%%%%%%%%%%%%%%%%%%%%%%%%%%%%
\bea
\label{FH5}
4x(x+1)\frac{d^2 S}{d x^2}+2(2x+1)\frac{d S}{d x}
+\left[
\tau-2\beta^2x(x+1)+2\gamma^2x+\frac{2\alpha^2}{x+1}+\frac{2 E}{x}
\right] S = 0.
\end{eqnarray}
%%%%%%%%%%%%%%%%%%%%%%%%%%%%%%%%%%
Choosing
%%%%%%%%%%%%%%%%%%%%%%%%%%%%%%%%%%
\bea
\label{FH6}
P(x)=(1+x)^{s}x^{t}e^{-\beta x /\sqrt{2}}
\prod\limits_{i=1}^{N}(x-\theta_i)
\end{eqnarray}
%%%%%%%%%%%%%%%%%%%%%%%%%%%%%%%%%%
where $t$ and $s$ are given by the formulas (\ref{FEP5}),
we obtain the energy spectrum (\ref{FEP9}).
Here $\theta_i$ satisfies the equations
%%%%%%%%%%%%%%%%%%%%%%%%%%%%%%%%%%
\bea
2\theta_i(1+\theta_i)\left(\sum^{N}_{\scriptstyle k=1 \atop\scriptstyle k\not=i}
\frac{1}{\theta_i-\theta_k}
-\frac{\beta}{\sqrt{2}}\right)
-2(1+\theta_i)N+
\frac{\sqrt{2}\gamma^2}{4\beta}\theta_i
&+&
\nonumber
\\[2mm]
\label{FH7}
\frac{\gamma^2}{\sqrt{2}\beta}-
\sqrt{2\alpha^2+\frac{1}{4}}-1&=&0.
\end{eqnarray}
%%%%%%%%%%%%%%%%%%%%%%%%%%%%%%%%%%
The separation constant $\tau$ is:
%%%%%%%%%%%%%%%%%%%%%%%%%%%%%%%%%%
\bea
\label{FH10}
\tau =
\frac{8\beta}{ \sqrt{2}} \sum_{i=1}^{N}\theta_i
-\left(\frac{\gamma^2}{\sqrt{2}\beta}-1\right)^2-\frac{4\beta}{\sqrt{2}}
\left(1+\sqrt{2 \alpha^2+\frac{1}{4}}\right)+2\gamma^2,
\end{eqnarray}
%%%%%%%%%%%%%%%%%%%%%%%%%%%%%%%%%%
so the total solution $\Psi( b, \theta)$ is represented as:
%%%%%%%%%%%%%%%%%%%%%%%%%%%%%%%%%%
\bea
\Psi_{Nlk} (b, \theta)&=& S_{Nl}(b)S_{Nk}(\theta)=
(\cosh b\cos\theta)^{\frac12+\sqrt{2\alpha^2+\frac14}}
(\sinh b\sin\theta)^{\frac{\gamma^2}{\sqrt{2}\beta}
-\sqrt{2\alpha^2+\frac14}-2N-\frac32}
\nonumber
\\[2mm]
\label{FH11}
&\cdot&
\exp\left\{-\frac{\beta}{\sqrt{2}} (\sinh^2b-\sin^2\theta)\right\}
\prod_{i=1}^{N}(\sinh^2 b-\theta_i)
(\sin^2\theta+\theta_i).
\end{eqnarray}
%%%%%%%%%%%%%%%%%%%%%%%%%%%%%%%%%%
The total number of zeros is $N$, and $k$ of them are located
in the interval $[-1,0]$  and $l$ are in $[0,\infty]$.

Each solution $\Psi_{Nlk}( b, \theta)$ satisfies the eigenvalue equation
%%%%%%%%%%%%%%%%%%%%%%%%%%%%%%%%%%
\begin{eqnarray}
\label{FHL}
L_{4} \Psi_{Nlk}(b, \theta)
&=&
-\frac{1}{\sin^2 b+\sin^2 \theta}\bigg\{
\sinh^2 b
\frac{\partial^2}{\partial b^2}
-\sin^2\theta
\frac{\partial^2}{\partial \theta^2}
-2\beta^2 (\cosh^2 b\sinh^4 b
\nonumber
\\[2mm]
&-&\cos^2 \theta\sin^4 \theta)
+2\gamma^2(\sinh^4 b-\sin^4\theta)+2\alpha^2(\tanh^2b+\tan^2\theta)
\bigg\}\Psi_{Nlk}(b, \theta)
\nonumber
\\[2mm]
&=&\bigg\{(K_2-M_1)^2-K_3^2+2\beta^2
\frac{(w_0+w_1)^2-w_2^2}
{(w_0-w_1)^2}-2\alpha^2\left(\frac{w_0-w_1}{w_2}\right)^2
\nonumber
\\[2mm]
&-&4\gamma^2\frac{w_1}{w_0-w_1}
\bigg\}\Psi_{Nlk}(b, \theta)
= \tau \Psi_{Nlk}(b, \theta).
\end{eqnarray}
%%%%%%%%%%%%%%%%%%%%%%%%%%%%%%%%%%

\subsection{Algebra}
Among the operators $\{L_1, L_2, L_3, L_4\}$, corresponding to the
four separable coordinate systems,
only two are independent, as
%%%%%%%%%%%%%%%%%%%%%%%%%%%%%%%%%%
\begin{eqnarray}
\label{AL0}
L_3=-L_2-L_1,\quad
L_4=L_2-L_1.
\end{eqnarray}
%%%%%%%%%%%%%%%%%%%%%%%%%%%%%%%%%%
Consider the operators  $N_1$, $N_2$ and $R$ where
%%%%%%%%%%%%%%%%%%%%%%%%%%%%%%%%%%
\begin{eqnarray}
\label{AL1}
N_1 &=& L_1,  \,\,\,\,\,\,\,
N_2 = L_2 - 2\gamma^2,
\nonumber\\[2mm]
R &\equiv& [N_1, N_2] =
2\{K_3,\{K_2,M_1\}\}-2 \{K_3,K_2^2\} -2 \{K_3,M_1^2\}
+8\bigg[ \alpha^2\bigg(\frac{\omega_0-\omega_1}{\omega_2}\bigg)^2+
\nonumber
\\[2mm]
&+&\beta^2\bigg(\frac{\omega_2}{\omega_0-\omega_1}\bigg)^2\bigg] K_3
+\frac{16\beta^2\omega_2}{(\omega_0-\omega_1)^2}(\omega_0 K_2-\omega_1 M_1)
+\frac{8\gamma^2\omega_2}{\omega_0-\omega_1}(M_1-K_2)
\nonumber
\\[2mm]
&-&4\left[\gamma^2+
2 \alpha^2\bigg(\frac{\omega_0-\omega_1}{\omega_2}\bigg)^2
- 2\beta^2\frac{1+2\omega_2^2}{(\omega_0-\omega_1)^2}\right]
\end{eqnarray}
%%%%%%%%%%%%%%%%%%%%%%%%%%%%%%%%%%
We have
%%%%%%%%%%%%%%%%%%%%%%%%%%%%%%%%%%
\begin{eqnarray}
[ R, N_2] &=& 8 N_2^2 + 64 \beta^2 H + 16 \gamma^2 N_2
+32 \beta^2 N_1 + 16\beta^2(1 - 4\alpha^2)
\\[2mm]
[ R, N_1] &=& -8 \{ N_1, N_2\} - 32\gamma^2 H + 16N_2 -16\gamma^2 N_1 +
16 \gamma^2(1-2\alpha^2)
\\[2mm]
R^2 &=& \frac{8}{3}\{N_2, N_2, N_1\} - \frac{176}{3}N_2^2 + 32\beta^2 N_1^2
+ 128 \beta^2 H^2 + 64\gamma^2 H N_2 + 128\beta^2 H N_1
\nonumber\\[2mm]
&+&
16\gamma^2 \{N_1, N_2\} +
\left(\frac{128}{3} + 256 \alpha^2\right)
\beta^2 H +
\left(64\alpha^2\gamma^2 - \frac{352}{3}\gamma^2\right) N_2
\nonumber\\[2mm]
&+& \left(\frac{352}{3} -128 \alpha^2\right)
\beta^2 N_1 +
(128\alpha^4 \beta^2 + 128\gamma^4\alpha^2 - \frac{128}{3}\alpha^2\beta^2
- \frac{64}{3}\beta^2 - 48\gamma^4)
\nonumber
\end{eqnarray}
%%%%%%%%%%%%%%%%%%%%%%%%%%%%%%%%%%
where $\{A,B\} = AB+BA$ and
$$
\{A,B,C\} = ABC + ACB + BCA + BAC + CAB + CBA.
$$
The integrals of motion $N_1, N_2$ and $H$ generate a quadratic algebra.

\subsection{Interbasis expansion}

For a fixed value of energy, we can write the
equidistant wave function (\ref{FE2}) in terms of the horicyclic ones
(\ref{FHC}) as
%%%%%%%%%%%%%%%%%%%%%%%%%%%%%%%%%%%%%%%%%%%%%%%%%%%%%%%%%%%%
\begin{equation}
\label{IB1}
\Psi_{n_1 n_2} (x,y) =
\sum_{m=0}^{n_1+n_2} W_{n_1 n_2}^{n m}(\alpha, \beta, \gamma)
\Psi_{n m} (a,b)
\end{equation}
%%%%%%%%%%%%%%%%%%%%%%%%%%%%%%%%%%%%%%%%%%%%%%%%%%%%%%%%%%%%
where $n_1 + n_2 = n + m$. The connection between the equidistant $(a,b)$
and horicyclic $(x, y)$ coordinates is
%%%%%%%%%%%%%%%%%%%%%%%%%%%%%%%%%%%%%%%%%%%%%%%%%%%%%%%%%%%%
\begin{equation}
\label{IB2}
x=e^b \tanh a ,\quad
y=e^b \frac{1}{\cosh a}.
\end{equation}
%%%%%%%%%%%%%%%%%%%%%%%%%%%%%%%%%%%%%%%%%%%%%%%%%%%%%%%%%%%%
Going over to the horicyclic coordinates in the left side of
expansion (\ref{IB1}), then considering the limit
$b \rightarrow\infty$ and using the
asymptotic formula for Laguerre polynomials \cite{EMOT}
%%%%%%%%%%%%%%%%%%%%%%%%%%%%%%%%%%%%%%%%%%%%%%%%%%%%%%%%%%%%
\begin{equation}
\label{IB3}
\lim_{x\rightarrow\infty}  L_n^{\alpha} (x) \rightarrow
(-1)^n\frac{x^n}{n!}
\end{equation}
%%%%%%%%%%%%%%%%%%%%%%%%%%%%%%%%%%%%%%%%%%%%%%%%%%%%%%%%%%%%
we see that dependence on $b$ cancels on both sides of (\ref{IB1}).
Now using the orthogonality condition for the angular wave functions
(\ref{FE7}) we find the following expression for the interbasis
coefficients $W_{n_1 n_2}^{n m}$:
%%%%%%%%%%%%%%%%%%%%%%%%%%%%%%%%%%%%%%%%%%%%%%%%%%%%%%%%%%%%
\begin{eqnarray}
\label{IB4}
W_{n_1 n_2}^{n m} = (-1)^{n}
\sqrt{\frac{ m! n! \sqrt{2} \beta (\mu-d-2n-1)\Gamma(\mu+m+1)\Gamma(\mu-n)}
{n_1!n_2! \mu \Gamma(n_1+d+1)\Gamma(n_2+d+1)\Gamma(n+d+1)\Gamma(\mu-d-n) }}
B^{n m}_{n_1 n_2}
\end{eqnarray}
%%%%%%%%%%%%%%%%%%%%%%%%%%%%%%%%%%%%%%%%%%%%%%%%%%%%%%%%%%%%
where
%%%%%%%%%%%%%%%%%%%%%%%%%%%%%%%%%%%%%%%%%%%%%%%%%%%%%%%%%%%%
\begin{eqnarray}
\label{IB5}
B_{n_1 n_2}^{n m} &=& \int_{-\infty}^{+\infty}
(\sinh a)^{1+2d +2 n_1} (\cosh a)^{1-2\mu-2 m}
P_n^{(d,-\mu)}(\cosh 2 a) \,\,da
\end{eqnarray}
%%%%%%%%%%%%%%%%%%%%%%%%%%%%%%%%%%%%%%%%%%%%%%%%%%%%%%%%%%%%
and $d=\sqrt{2 \alpha^2+1/4}$.
The integral $B_{n_1, n_2}^{n m}$ can be evaluated by expressing the Jacobi
polynomial through the hypergeometric function ${_2F_1}$ \cite{EMOT}:
%%%%%%%%%%%%%%%%%%%%%%%%%%%%%%%%%%%%%%%%%%%%%%%%%%%%%%%%%%%%
\begin{eqnarray}
\label{IB6}
P_n^{(\alpha, \beta)}(x)=(-1)^n\frac{\Gamma(n+\beta+1)}{\Gamma(\beta+1) n!}
{_2F_1} \left(\left.\matrix{-n,\, n+\alpha+\beta+1\cr
\beta+1\, \cr}
\right|\frac{1+x}{2}\right).
\end{eqnarray}
%%%%%%%%%%%%%%%%%%%%%%%%%%%%%%%%%%%%%%%%%%%%%%%%%%%%%%%%%%%%
Representing the function $_2F_1$ as a series we
come to a sum of  integrals, each of  which can be calculated
by using the formula \cite{EMOT}:
%%%%%%%%%%%%%%%%%%%%%%%%%%%%%%%%%%%%%%%%%%%%%%%%%%%%%%%%%%%%
\begin{eqnarray}
\label{IB7}
\int_{0}^{+\infty}
(\sinh \tau)^{\alpha} (\cosh \tau)^{-\beta}\,\,d \tau=\frac{1}{2}
B\left({\frac{1+\alpha}{2},\frac{\beta-\alpha}{2}}\right),\quad
[Re \alpha>-1, Re(\alpha-\beta)<0].
\end{eqnarray}
%%%%%%%%%%%%%%%%%%%%%%%%%%%%%%%%%%%%%%%%%%%%%%%%%%%%%%%%%%%%
We thus obtain
%%%%%%%%%%%%%%%%%%%%%%%%%%%%%%%%%%%%%%%%%%%%%%%%%%%%%%%%%%%%
\begin{eqnarray}
\label{IB8}
W_{n_1 n_2}^{n m} =
\frac{(-1)^n}{2}\sqrt{\frac{m!\sqrt{2}\beta(\mu-d-2 n-1)
(\mu+m)\Gamma(n_1+d+1)}
{n! n_1! n_2!\mu \Gamma(n_2+d+1) \Gamma(n+d+1) \Gamma(\mu-n-d)}}
\\[2mm]
\frac{\Gamma(\mu) \Gamma(\mu+m-d-n_1-1)}
{\sqrt{\Gamma(\mu-n) \Gamma(\mu+m)}}
{_3F_2} \left(\left.\matrix{-n,\, n+d-\mu+1, \, 1-\mu-m\cr
1-\mu, \, 2+n_1+d-\mu-m \cr}
\right|1\right).
\nonumber
\end{eqnarray}
%=============================================================
Alternatively, by using the formula
\cite{SUS} for the Hahn polynomials $h_n^{(\alpha,\beta)}(x,N)$.
%=============================================================
\begin{eqnarray}
\label{IB9}
h_n^{(\alpha,\beta)}(x,N)=
\frac{(-1)^n\Gamma(N)\Gamma(\beta+n+1)}
{n!\Gamma(N-n)\Gamma(\beta+1)}
{}_3 F_2\biggl(\matrix{-n \mbox{  }  ;
\mbox{  } \alpha+\beta+n+1 \mbox{  }  ;
\mbox{  }-x\hfill\cr
\beta+1 \mbox{    }  ;\mbox{    }  1-N\hfill\cr}
\bigg\vert 1 \biggr)
\end{eqnarray}
%=============================================================
we obtain the following expression for the expansion coefficients
%=============================================================
\begin{eqnarray}
\label{IB10}
W_{n_1 n_2}^{n m} &=&
\frac{(-1)^n}{2}\sqrt{\frac{m!n!\sqrt{2}\beta(\mu-d-2 n-1) (\mu+m)}
{ n_1! n_2!\mu \Gamma(n+d+1) \Gamma(\mu-n-d)}}
\nonumber
\\[2mm]
&\cdot&\sqrt{\frac{\Gamma(n_1+d+1)\Gamma(\mu-n)}
{\Gamma(n_2+d+1)\Gamma(\mu+m)}}
\Gamma(\mu+m-d-n_1-n-1)
\\[2mm]
&\cdot&h_n^{( d , -\mu)} (\mu+m+1, \mu+m-d-n_1-1),
\nonumber
\end{eqnarray}
%%%%%%%%%%%%%%%%%%%%%%%%%%%%%%%%%%%%%%%%%%%%%%%%%%%%%%%%%%%%
in terms of Hahn polynomials.

\hspace{0.5cm}

\section{Second Potential.}

The second considered potential is
%=============================================================
\begin{equation}
\label{S1}
V_2 = \frac{\alpha^2}{\omega_2^2}
+\gamma^2\frac{\omega_0\omega_1}{(\omega_0^2+\omega_1^2)^2}
+(\alpha^2-\beta^2)\frac{\omega_0^2-\omega_1^2}{(\omega_0^2+\omega_1^2)^2}
\end{equation}
%=============================================================
where $\alpha, \beta$ and $\gamma$  are positive constants.
The corresponding Schr\"odinger equation admits separable solutions in two
coordinate systems: equidistant and semi--hyperbolic.

\subsection{Solutions of the Schr\"odinger equation}

\vspace{0.5cm} \underline{{\it 2.1 Equidistant coordinates.}}
In this coordinate system
%=============================================================
\bea
\label{SE0}
\omega_0 = \cosh \tau_1 \cosh \tau_2,
\,\,\,\,
\omega_1 = \cosh \tau_1 \sinh \tau_2,
\,\,\,\,
\omega_2 = \sinh \tau_1,
\eea
%=============================================================
$[\tau_1,\tau_2 \in (-\infty, \infty)]$,
the potential $V_2$ has the form
%=============================================================
\begin{equation}
\label{SE1}
V_2(\tau_1,\tau_2) =
-\frac{\alpha^2}{\sinh^2 \tau_1}
+\frac{1}{\cosh^2 \tau_1}
\frac{\alpha^2-\beta^2+\gamma^2\cosh \tau_2\sinh \tau_2}
{(\cosh^2 \tau_2+\sinh^2 \tau_2)^2}.
\end{equation}
%=============================================================
After putting
%=============================================================
\begin{equation}
\label{SE2}
\Psi (\tau_1,\tau_2) = (\cosh \tau_1)^{-1/2} S(\tau_1) Z(\tau_2)
\end{equation}
%=============================================================
we arrive at two  equations:
%=============================================================
\begin{eqnarray}
\label{SE3}
\frac{d^2 S}{d \tau_2^2} &+& \left[ -\mu^2-
\frac{2(\alpha^2-\beta^2) +
\gamma^2\sinh (2\tau_2)}{\cosh^2(2\tau_2)}\right] S = 0
\\[2mm]
\label{SE4}
\frac{d^2 S}{d \tau_1^2} &+& \left[2 E-\frac{1}{4} +
\frac{\mu^2-\frac{1}{4}}{\cosh^2\tau_1}
- \frac{2\alpha^2}{\sinh^2 \tau_1} \right] S = 0
\end{eqnarray}
%=============================================================
where $\mu$ is the equidistant separation constant.

Let us consider the first equation (\ref{SE3}).
The substituon $x=\sinh2\tau_2$ transforms this equation to
%=============================================================
\begin{eqnarray}
\label{SE5}
4(1+x^2)\frac{d^2 S}{d x^2}+4x
\frac{d S}{d x}+\left[-\mu ^2+{2(\beta ^2-\alpha ^2)
-\gamma^2x \over (1+x^2)}\right] S = 0
\end{eqnarray}
%=============================================================
where the physical region is $x\in(-\infty, \infty)$.
The equation (\ref{SE3}) has three regular singularities
in the point $x = -i, i, \infty$ and may be solved in term
of hypergeometric functions. The solution of the equation
(\ref{SE3}) for a large $|x|$ can be written as:
%=============================================================
$$
S(x)
= A_1 \,
(x-i)^{-\frac{b+\mu}{2}-\frac{1}{4}}
(x+i)^{\frac{b}{2}+\frac{1}{4}}
\,_2F_1\left(\frac{a+b+1+\mu}{2},\frac{b-a+1+\mu}{2};
\, 1+ \mu;\,
\frac{2i}{i-x}\right)
$$
\begin{eqnarray}
\label{SE6}
+ \, A_2 \,
(x-i)^{-\frac{b-\mu}{2}-\frac{1}{4}}
(x+i)^{\frac{b}{2}+\frac{1}{4}}
\,_2F_1\left(\frac{a+b+1-\mu}{2},\frac{b-a+1-\mu}{2};
\, 1-\mu; \,
\frac{2i}{i-x}\right)
\end{eqnarray}
%=============================================================
with
%=============================================================
\begin{eqnarray}
\label{SE7}
a^2=(b^2)^*=
\frac{{2\beta^2-2\alpha^2+1-i\gamma^2}}{4}.
\end{eqnarray}
%=============================================================

Let the separation constant $\mu$  be a positive number
(the equation (\ref{SE6}) is symmetric with respect to the
replacement $\mu\to -\mu$). Then the second term in formula
(\ref{SE6}) behaves like $|x|^{\frac{\mu}{2}}$ at $\infty$
and must be omitted. Thus for $S(x)$ we obtain
%=============================================================
\begin{eqnarray}
\label{SE8}
S(x) =
A \, (x-i)^{-\frac{b+\mu}{2}-\frac{1}{4}}
(x+i)^{\frac{b}{2}+\frac{1}{4}}
\,_2F_1\left(\frac{a+b+1+\mu}{2},\frac{b-a+1+\mu}{2}; \mu+1;
\frac{2i}{i-x}\right)
\end{eqnarray}
%=============================================================
\begin{figure}[htb]
\centerline{\epsfxsize=4in\epsfbox{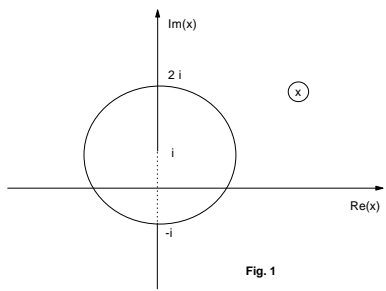}}
\caption{Domain of Convergence}\label{fig1}
\end{figure}
%============================================================
The hypergeometric function in equation (\ref{SE8}) converges
if $x$ lies out of the circle $C$ on  Fig.1, defined by
$|i-x| = 2$, and converges on the circle $C$ with the condition
$Re(b)<0$. The function $S(x)$ exists everywhere inside $C$
except the interval $x \in [-i, i]$ (see \ref{fig1}), since the
hypergeometric function in (\ref{SE8}) has a cut along the
argument $\frac{2 i}{i-x}\in [1,\infty)$). That means that
the solution (\ref{SE8}) along the real axes inside $C$ in
general is not a continuous function and may have a jump at
the point $x=0$. Let us now consider the analytic continuation
of (\ref{SE8}) inside the circle $C$

%=============================================================
\begin{eqnarray}
S(x) &=& A \bigg\{
(x-i)^{\frac{a}{2}+\frac{1}{4}}(x+i)^{\frac{b}{2}+\frac{1}{4}}
\frac{\Gamma(\mu+1)\Gamma(-a)}{\Gamma(\frac{b-a+1+\mu}{2})
\Gamma(\frac{-b-a+1+\mu}{2})(2 i)^{\frac{a+b+1+\mu}{2}}}
\nonumber
\\[2mm]
&\cdot&\,_2F_1\left(\frac{a+b+1+\mu}{2},\frac{a+b+1-\mu}{2};
a+1; \frac{i-x}{2i}\right)
\nonumber
\\[2mm]
&+&
(x-i)^{-\frac{a}{2}+\frac{1}{4}}
(x+i)^{\frac{b}{2}+\frac{1}{4}}
\frac{\Gamma(\mu+1)\Gamma(a)}{\Gamma(\frac{b+a+1+\mu}{2})
\Gamma(\frac{-b+a+1+\mu}{2})(2 i)^{\frac{-a+b+1+\mu}{2}}}
\nonumber
\\[2mm]
\label{SE9}
&\cdot&\,
_2F_1\left(\frac{-a+b+1+\mu}{2},\frac{-a+b+1-\mu}{2}; -a+1;
\frac{i-x}{2i}\right)\bigg\}.
\end{eqnarray}
%=============================================================
>From equation (\ref{SE7}) follow two possibilities
%=============================================================
\begin{eqnarray}
a = b^*  , \qquad a = - b^*.
\end{eqnarray}
%=============================================================
Putting the $a=b^*$ ($Re(a) = Re(b)<0$) we find that the first
term in (\ref{SE9}) represents an analytic function,
while the second term is discontinuous at $x=0$.
[Note since the both terms in equation (\ref{SE9}) transform
to each other with replacement $a\rightarrow -a$ the choice
$a= - b^*$ means that the first term in (\ref{SE9}) is
discontinuous while the second term is continous at $x=0$.]
Thus the {\it sufficient} condition for the existence of the
continuous solution requires the relation
%=============================================================
\begin{eqnarray}
\label{SE10}
\mu + a + a^* + 1 =- 2m, \qquad m=0,1,2...,
\left[-\frac{a+a^* +1}{2}\right],
\end{eqnarray}
%=============================================================
so from (\ref{SE7}) we have
%=============================================================
\begin{eqnarray}
\label{SE11}
\mu =
- 2m - 1+\frac{1}{\sqrt{2}}
\sqrt{2\beta^2-2\alpha^2+1+\sqrt{(2\beta^2-2\alpha^2+1)^2+\gamma^4}}.
\end{eqnarray}
%=============================================================
Finally, the orthonormalized eigenfunction of equation
(\ref{SE3}) may be written in the form
%=============================================================
\begin{eqnarray}
S(\tau_2) &=&
(-1)^{\frac{3m}{2}}
\Gamma(-a)
\sqrt{\frac{(-2m-a-a^*-1)\Gamma(-m-a^*)}
{\pi m! 2^{a+a^*+1} \Gamma(-m-a)\Gamma(-m-a-a^*)}}
\nonumber
\\[2mm]
&\cdot&
(1+i \sinh 2\tau_2)^{\frac{a}{2}+\frac{1}{4}}
(1-i \sinh 2\tau_2)^{\frac{a^*}{2}+\frac{1}{4}}
\nonumber
\\[2mm]
&\cdot&
{}_2F_1\left(-m,m+a+a^*+1; a+1;
\frac{1+i\sinh 2\tau_2}{2}\right)
\nonumber
\\[2mm]
&=&
(-1)^{\frac{m}{2}}
\sqrt{\frac{
(-2m-a-a^*-1)m!\Gamma(-m-a)\Gamma(-m-a^*)}
{\pi 2^{a+a^*+1}\Gamma(-m-a-a^*)}}
\nonumber
\\[2mm]
&\cdot&
\label{SE13}
(1+i\sinh 2\tau_2)^{\frac{a}{2}+\frac{1}{4}}
(1-i \sinh 2\tau_2)^{\frac{a^*}{2}+\frac{1}{4}}
P_m^{(a,a^*)}(-i \sinh 2\tau_2)
\end{eqnarray}
%=============================================================
where
%=============================================================
\begin{eqnarray*}
a=
\frac{1}{2^\frac32}\Bigg\{
-\sqrt{\sqrt{(2\beta^2-2\alpha^2+1)^2+\gamma^4}+ 2\beta^2-2\alpha^2+1}
\\[2mm]
+i\,\,\sqrt{\sqrt{(2\beta^2-2\alpha^2+1)^2+\gamma^4}-(2\beta^2-2\alpha^2+1)}
\Bigg\}.
\end{eqnarray*}
%=============================================================
The second equation (\ref{SE4}) is quite like  (\ref{FE4})
and has a solution:
%=============================================================
\begin{eqnarray}
\label{SE15}
Z(\tau_1) &\equiv& S^{(\alpha, \mu)}_{n} (\tau_1)
= \sqrt{\frac{2 (\mu-\sqrt{2\alpha^2+1/4}-2n-1)\Gamma(\mu-n)n!}
{\Gamma(\mu-\sqrt{2\alpha^2+1/4}-n)\Gamma(1+n+\sqrt{2\alpha^2+1/4})}}
\nonumber
\\[2mm]
&\times&
(\sinh \tau_1)^{\frac{1}{2}+\sqrt{2\alpha^2+1/4}}
(\cosh \tau_1)^{\frac{1}{2}-\mu}
P^{(\alpha, -\mu)}_n(\cosh 2 \tau_1)
\end{eqnarray}
%=============================================================
with $n= 0,1, ...$.

The quantized energy is
%=============================================================
\bea
E = -\frac{1}{2}(\mu-\sqrt{2\alpha^2+1/4}-2n-1)^2+\frac18
=
-\frac{1}{2}\Bigg\{2 N+2+\sqrt{2\alpha^2+1/4}
\nonumber
\\[2mm]
\label{SE16}
-
\frac{1}{\sqrt{2}}
\sqrt{2\beta^2-2\alpha^2+1+\sqrt{(2\beta^2-2\alpha^2+1)^2+\gamma^4}}
\Bigg\}^2+\frac18
\eea
%=============================================================
where $N=n+m$ is the principal quantum number and the bound state
occurs for
%=============================================================
\bea
0 \leq N \leq \left[\frac{1}{\sqrt{8}}
\sqrt{2\beta^2-2\alpha^2+1+\sqrt{(2\beta^2-2\alpha^2+1)^2+\gamma^4}}
- \frac{1}{2}\sqrt{2\alpha^2+1/4} - 1\right].
\eea
%=============================================================

The additional operator describing this coordinate system is
%=============================================================
\begin{eqnarray}
L_1 \Psi_{nm}(\tau_1, \tau_2)\equiv
\left[K_3^2
-2(\alpha^2-\beta^2)
\left(\frac{\omega_0^2-\omega_1^2}{\omega_0^2+\omega_1^2}\right)^2
-2\gamma^2\frac{\omega_0\omega_1
(\omega_0^2-\omega_1^2)}{(\omega_0^2+\omega_1^2)^2}
\right] \Psi_{nm}(\tau_1, \tau_2)
\nonumber
\\[2mm]
\label{SE17}
=
\left\{2m+1
- \frac{1}{\sqrt{2}}
\sqrt{2\beta^2-2\alpha^2+1+\sqrt{(2\beta^2-2\alpha^2+1)^2+\gamma^4}}
\right\}^2
\Psi_{nm}(\tau_1, \tau_2).
\end{eqnarray}
%=============================================================

\vspace{0.5cm}
\underline{{\it 2.2 Semi--hyperbolic coordinates}}. Here
%=============================================================
\bea
\label{SSH0}
\omega_0^2 &=&-\frac{(\mu-e_3)(\nu-e_3)}{2[(e_3-a)^2+b^2]}+\frac{1}{2}
-\frac{1}{2 b}
\bigg[\frac{[(\mu-a)^2+b^2][(\nu-a)^2+b^2]}{(e_3-a)^2+b^2}\bigg]^{1/2}
\nonumber
\\[2mm]
\omega_1^2 &=&\frac{(\mu-e_3)(\nu-e_3)}{2[(e_3-a)^2+b^2]}-\frac{1}{2}
-\frac{1}{2 b}
\bigg[\frac{[(\mu-a)^2+b^2][(\nu-a)^2+b^2]}{(e_3-a)^2+b^2}\bigg]^{1/2}
\\[2mm]
\omega_2^2 &=&-\frac{(\mu-e_3)(\nu-e_3)}{(e_3-a)^2+b^2}
\nonumber
\eea
%=============================================================
$[\nu<e_3<\mu]$, where $\sinh 2f = (e_3-a)/b$ and $2 f$
is the distance between the focii of the semi-hyperbolas
and the bases of their equidistants \cite{OLE}.

If we change  variables according to
%=============================================================
\begin{equation}
\label{SSH1}
\omega _0=(s_1+s_2)/\sqrt 2,\quad
\omega _1=-i(s_1-s_2)/\sqrt 2,\quad
\omega _2=-is_3,
\end{equation}
%=============================================================
the Schr\"odinger equation becomes
%=============================================================
\begin{eqnarray}
\frac12\bigg[\bigg
(s_1\frac{\partial}{\partial s_2}
-s_2\frac{\partial}{\partial s_1}\bigg)^2+
\bigg(s_1\frac{\partial}{\partial s_3}
-s_3\frac{\partial}{\partial s_1}\bigg)^2 +
\bigg(s_3\frac{\partial}{\partial s_2}
-s_2\frac{\partial}{\partial s_3}\bigg)^2
\bigg]\Psi
\nonumber
\\[2mm]
\label{SSH2}
\bigg[- E - \frac12\left({k^2_1-{1\over 4}\over s^2_1}
+ {k^2_2-{1\over 4}\over s^2_2} +
{k^2_3-{1\over 4}\over s^2_3}\right)\bigg]\Psi
=0
\end{eqnarray}
%=============================================================
with
%=============================================================
\begin{eqnarray}
\label{SSH3}
{1\over 2}(k^2_1-{1\over 4})={1\over 4}(\beta ^2-\alpha ^2)-
{i\over 8}\gamma ^2,\quad
{1\over 2}(k^2_2-{1\over 4})={1\over 4}(\beta ^2-\alpha ^2)
+{i\over 8}\gamma ^2,\quad
{1\over 2}(k^2_3-{1\over 4})=\alpha ^2.
\nonumber
\end{eqnarray}
%=============================================================
Noting
%=============================================================
$$
\omega ^2_0-\omega ^2_1-\omega ^2_2=s^2_1+s^2_2+s^2_3=1
$$
%=============================================================
and considering eq.(\ref{SSH2}) we
see that the problem we wish to solve using the
real coordinates $\omega _0,\omega _1$ and $\omega _2$ is a real case of the
corresponding problem on the sphere with coordinates $s_1,s_2, s_3$
and  energy $\varepsilon=-E$.

Inverting the relations (\ref{SSH1}) we have
%=============================================================
\begin{eqnarray}
\label{SSH4}
s_1=(\omega _0+i\omega _1)/\sqrt 2,\quad
s_2=(\omega _0-i\omega _1)/\sqrt 2,\quad
s_3=i\omega _2.
\nonumber
\end{eqnarray}
%=============================================================
Now  choose elliptic coordinates on the complex sphere according to
%=============================================================
\begin{eqnarray}
\label{SSH5}
s^2_1={(\mu -e_1)(\nu -e_1)\over (e_1-e_2)(e_1-e_3)},\quad
s^2_2={(\mu -e_2)(\nu -e_2)\over (e_2-e_1)(e_2-e_3)},\quad
s^2_3={(\mu -e_3)(\nu -e_3)\over (e_3-e_2)(e_3-e_1)}.
\nonumber
\end{eqnarray}
%=============================================================
This choice of real coordinates $\mu, \nu$ will work for the real coordinates
$\omega _k,k=0,1,2$ if we take
$e_1=e^*_2=a+ib ,a,b$ real
and $\nu <e_3<\mu $.

In terms of the coordinates $\mu $ and $\nu $
the Schr\"odinger equation has the form:
%=============================================================
\begin{eqnarray}
\label{SSH6}
{4\over (\mu -\nu )}\bigg\{(\mu -e^*_2)(\mu -e_2)(\mu -e_3)
\bigg[{\partial ^2\Psi\over \partial \mu ^2}
+ {1\over 2}\bigg({1\over \mu -e^*_2} + {1\over \mu -e_2} +
{1\over \mu -e_3} \bigg){\partial \Psi \over \partial \mu }\bigg]
\nonumber
\\[2mm]
-(\nu-e^*_2)(\nu -e_2)(\nu -e_3)
\bigg[{\partial ^2\Psi \over \partial \nu ^2} -
{1\over 2}\bigg({1\over \nu -e^*_2} + {1\over \nu -e_2} +
{1\over \nu -e_3}\bigg){\partial \Psi \over \partial \nu }\bigg]\bigg\}
\nonumber
\\[2mm]
+ \bigg[(k^2_1-{1\over 4})
{(e^*_2-e_2)(e^*_2-e_3)\over (\mu  -e^*_2)(\nu  -e^*_2)} +
(k^2_2-{1\over 4}) {(e_2-e^*_2)(e_2-e_3)\over (\mu  -e_2)(\nu  -e_2)}
\nonumber
\\[2mm]
+ (k^2_3-{1\over 4}) {(e_3-e_2)(e_3-e^*_2)\over (\mu  -e_3)(\nu  -e_3)}\bigg]\Psi
=-2E\Psi.
\end{eqnarray}
%=============================================================
The separation equations are:
%=============================================================
\begin{eqnarray}
(\rho -e^*_2)(\rho -e_2)(\rho -e_3)
\bigg[{d ^2\Psi \over d \rho ^2} +
{1\over 2}\bigg({1\over \rho -e^*_2} + {1\over \rho -e_2} +
{1\over \rho -e_3}\bigg){d \Psi \over d \rho }\bigg]
\nonumber
\\[2mm]
-\frac{1}{4}\bigg[(k^2_1 - {1\over 4}){(e^*_2-e_2)(e^*_2-e_3)\over
(\rho -e^*_2)} +(k^2_2 -{1\over 4}){(e_2-e^*_2)(e_2-e_3)\over (\rho  -e_2)}
\nonumber
\\[2mm]
\label{SSH7}
+(k^2_3 - {1\over 4}){(e_3-e_2)(e_3-e^*_2)\over (\rho  -e_3)}
-2E\rho +\lambda \bigg]\psi(\rho)  =0
\end{eqnarray}
%=============================================================
where $\rho  =\mu ,\nu $. The operator $L_2$ with eigenvalue $\lambda $ is
%=============================================================
\begin{eqnarray}
L_2\Psi  =
{-4\over (\mu -\nu )}\bigg\{\nu (\mu -e_1)(\mu -e_2)(\mu -e_3)
\bigg[{\partial ^2\Psi \over \partial \mu ^2} +
{1\over 2}\bigg({1\over \mu -e^*_2} + {1\over \mu -e_2} +
{1\over \mu -e_3}\bigg){\partial \Psi \over \partial \mu }\bigg]
\nonumber
\\[2mm]
-\mu[(\nu -e_1)(\nu -e_2)(\nu -e_3)
\bigg[{\partial ^2\Psi \over \partial \nu ^2}
+{1\over 2}\bigg({1\over \nu -e^*_2} + {1\over \nu -e_2} +
{1\over \nu -e_3}\bigg){\partial \Psi \over \partial \nu }\bigg]\bigg\}
\nonumber
\\[2mm]
-\bigg[(k^2_1 -{1\over 4})
{(e^*_2-e_2)(e^*_2-e_3)\over (\mu -e^*_2)(\nu -e^*_2)}(\mu +\nu -e^*_2) +
(k^2_2 -{1\over 4})
{(e_2-e^*_2)(e_2-e_3)\over (\mu-e_2)(\nu-e_2)}(\mu +\nu -e_2)
\nonumber
\\[2mm]
\label{SSH8}
+ (k^2_3 -{1\over 4})
{(e_3-e_2)(e_3-e^*_2)\over (\mu-e_3)(\nu-e_3)}(\mu +\nu -e_3)\bigg]\Psi.
\end{eqnarray}
%=============================================================
In order to find the bound state solutions of this system
in semi--hyperbolic systems we first observe the identity
%=============================================================
\begin{eqnarray}
{s^2_1\over \theta _j-e^*_2}+{s^2_2\over \theta _j-e_2}+{s^2_3\over \theta _j-%
e_3} &=&
{(\omega ^2_0-\omega ^2_1)(\theta _j-a)
-2\omega _0\omega _1b\over (\theta _j-a)^2+b^2}
-\frac{\omega_2^2}{\theta_j-e_3}
\nonumber
\\[2mm]
&=&
\label{SSH9}
{(\mu -\theta _j)(\nu -\theta _j)\over (\theta _j-e^*_2)(\theta _j-e_2)(%
\theta _j-e_3)}.
\end{eqnarray}
%=============================================================
If we then look for solutions of the form
%=============================================================
\begin{eqnarray}
\label{SSH10}
\Psi = \prod ^3_{\ell =1}s^{k_\ell + {1\over 2}}_\ell
\prod ^N_{j=1}
\bigg({s^2_1\over \theta _j-e^*_2}+{s^2_2\over \theta _j-e_2}
+{s^2_3\over \theta_j - e_3}\bigg),
\end{eqnarray}
%=============================================================
we see that the corresponding zeros satisfy the equations
%=============================================================
\begin{eqnarray}
\label{SSH11}
{k_1+1\over \theta _m-e^*_2} + {k_2+1\over \theta _m-e_2} +
{k_3+1\over \theta _m-e_3} + \sum _{j\neq m}^N {2\over (\theta _m-\theta _j)} =0.
\end{eqnarray}
%=============================================================
For the energy $E$ we have
%=============================================================
\begin{eqnarray}
\label{SSH12}
E = -{1\over 2}(2N+2+k_1+k_2+k_3)^2+ {1\over 8},
\end{eqnarray}
%=============================================================
which coincides with the formula (\ref{SE8}), note (\ref{SSH3}).
For the separation constant $\lambda$ we obtain:
%=============================================================
\begin{eqnarray}
\lambda &=& -2[k_1(e_2+e_3)+k_2(e^*_2+e_3)+k_3(e_2+e^*_2)]
-2[e_3k_1k_2+e_2k_1k_3+e^*_2k_2k_3]
\nonumber
\\[2mm]
\nonumber
& -& {3\over 2}(e^*_2+e_2+e_3)
-4e_2e_3(k_1+1)\sum ^q_{m=1} {1\over (\theta _m-e^*_2)}
-e^*_2e_3(k_2+1)\sum ^q_{m=1} {1\over (\theta _m-e_2)}
\\[2mm]
\label{SSH13}
&-&4e_2e^*_2(k_3+1)\sum ^q_{m=1} {1\over (\theta _m-e_3)}.
\end{eqnarray}
%=============================================================
In terms of variables $w_i$ the total wave function is written:
%=============================================================
\begin{eqnarray*}
\label{SSH14}
\Psi =\bigg(\frac{\omega _0+i\omega _1}{\sqrt 2}\bigg)^{k_1 + {1\over 2}}
\bigg(\frac{\omega _0-i\omega _1}{\sqrt 2}\bigg)^{k_2 + {1\over 2}}
(i\omega _2)^{k_3 + {1\over 2}}
\prod ^N_{j=1}
\bigg[{(\omega ^2_0-\omega ^2_1)(\theta _j-a)
-2\omega _0\omega _1b\over (\theta _j-a)^2+b^2}
-\frac{\omega_2^2}{\theta_j-e_3}\bigg].
\end{eqnarray*}
%=============================================================

The algebra of second order symmetries for this potential
is generated by the operators
%=============================================================
\begin{eqnarray}
\label{SSH15}
L_{jk}=(s_j\partial _{s_k}-s_k\partial _{s_j})^2+({1\over 4}-k^2_j){s^2_k%
\over s^2_j} +({1\over 4}-k^2_k){s^2_j\over s^2_k}
\end{eqnarray}
%=============================================================
for $j,k=1,2,3$ and $j\neq k$.
The Hamiltonian of the system is expressed in terms of $L_{jk}$ as:
%=============================================================
\begin{eqnarray}
\label{SSH15'}
H=\frac12(L_{12}+L_{13}+L_{23})-\frac12\sum_{i=1}^{3}k_i^2+\frac34.
\end{eqnarray}
%=============================================================
The relevant generators in the real case we are
considering are then
%=============================================================
\begin{eqnarray}
\label{SSH16}
L_{12} &=& -K_3^2+
\left(\frac{1}{4}-k_1^2\right)
\left(\frac{\omega_0-i \omega_1}{\omega_0+i \omega_1}\right)^2
+\left(\frac{1}{4}-k_2^2\right)
\left(\frac{\omega_0+i \omega_1}{\omega_0-i \omega_1}\right)^2
\\[2mm]
L_{13}&=&\frac12(M_1-iK_2)^2+
\left(\beta ^2-\alpha ^2-
{i\over 2}\gamma ^2\right)\frac{\omega_2^2}{(\omega_0+i\omega_1)^2}
+\alpha^2
\frac{(\omega_0+i\omega_1)^2}{w_2^2}
\\[2mm]
L_{23}&=&\frac12(M_1+iK_2)^2+
\left(\beta ^2-\alpha ^2+
{i\over 2}\gamma ^2\right)\frac{\omega_2^2}{(\omega_0-i\omega_1)^2}
+\alpha^2
\frac{(\omega_0-i\omega_1)^2}{w_2^2}.
\end{eqnarray}
%=============================================================
The commutation relations and resulting quadratic algebra can
then be deduced from the relations for the complex forms in
terms of the $L_{ij}$.
It is easy to show that the additional integrals of motion, corresponding to
the separation in
equidistant and semi--hyperbolic coordinates can be written as
%%%%%%%%%%%%%%%%%%%%%%%%%%%%%%%%%%%%%
\begin{eqnarray}
\label{SSH17}
L_1=-L_{12}+\beta^2-\alpha^2
\end{eqnarray}
%%%%%%%%%%%%%%%%%%%%%%%%%%%%%%%%%%%%%
and
%%%%%%%%%%%%%%%%%%%%%%%%%%%%%%%%%%%%%
\begin{eqnarray}
L_2&=&e_3 L_{12}+e_2 L_{13}+e_1 L_{32}
-k_1^2(e_2+e_3-e_1)
-k_2^2(e_1+e_3-e_2)
-k_3^2(e_1+e_2-e_3)
\nonumber
\\[2mm]
\label{SSH18}
&+&\frac{1}{4}(e_1+e_2+e_3).
\end{eqnarray}
%%%%%%%%%%%%%%%%%%%%%%%%%%%%%%%%%%%%%
The algebra for the operators (\ref{SSH17}), (\ref{SSH18}) is
found in the work \cite{KMJP}.

\section*{Acknowledgment(s)}
The authors thank professors V.M.Ter-Antonyan and
P.Winternitz for interesting discussions.

\vspace{1cm}

%%%%%%%%%%%%%%%%%%%%%%%%%%%%%%%%%%


\begin{thebibliography}{99}
%-----------------------------------------------------------------------
\bibitem{BKW}  C.P.Boyer, E.G.Kalnins and P.Winternitz.
Completely integrable relativistic Hamiltonian systems and
separation of variables in Hermitian hyperbolic spaces.
{\it J.Math.Phys.} {\bf 24}, 2022 (1983).
%-----------------------------------------------------------------------
\bibitem{GPSI}  C.Grosche, G.S.Pogosyan, A.N.Sissakian. Path Integral
Approach to Superintegrable Potentials. The Two-Dimensional Hyperboloid.
{\it Phys.Part.Nucl.} {\bf 27}, 244, (1996).
%-----------------------------------------------------------------------
\bibitem{KMJP1}  E.G.Kalnins, W.Miller Jr. and G.S.Pogosyan.
Superintegrability on the two dimensional hyperboloid;
{\it J.Math.Phys.} {\bf 38}, 5416, (1997).
%-----------------------------------------------------------------------
\bibitem{WLS} P.Winternitz, I.Lukac, and Ya.A.Smorodinskii. Quantum
numbers in the little groups of the Pioncare' group. {\it Sov.J.Nucl.Phys.} {\bf 7}, 139, 1968.
%-----------------------------------------------------------------------
\bibitem{KM} E.G.Kalnins and W.Miller Jr. Lie theory and separation of
variables, 4: The groups $SO(2,1)$ and $SO(3)$. {\it J.Math.Phys.}
{\bf 15}, 1263, 1974.
%-----------------------------------------------------------------------
\bibitem{OLE}  M.N.Olevski\u {i}. Triorthogonal Systems in Spaces of
Constant Curvature in which the Equation $\Delta _{2}u+\lambda u=0$ Allows
the Complete Separation of Variables {\it Math.Sb.}, {\bf 27}, 379 (1950)
(in russian).
%-----------------------------------------------------------------------
\bibitem{EMOT}  A.Erd\'{e}lyi, W.Magnus, F.Oberhettinger and F.G.Tricomi
(eds) Higher Transcendental Functions, Vol. I,II {\it McGraw Hill},\ New
York, 1953.
%-----------------------------------------------------------------------
\bibitem{FWO}  A.Frank and K.B.Wolf. Lie Algebras for Systems with Mixed
Spectra. I. The Scattering P\"{o}schl-Teller Potential. {\it J.Math.Phys.},
{\bf 25}, 973, (1985).
%-----------------------------------------------------------------------
\bibitem{LN}
L.D.Landau, E.M.Lifshitz, {\it Quantum Mechanics: Non-relativistic Theory},
Pergamon, Oxford, 1977.
%----------------------------------------------------------------------
\bibitem{SUS} A.F.Nikiforov, S.K.Suslov, V.B.Uvarov,
1985 {\it Classical orthogonal polynomials of discrete variables}, Nauka, Leningrad, 1985.
%----------------------------------------------------------------------
\bibitem{KMJP}  E.G.Kalnins, W.Miller Jr. and G.S.Pogosyan.
Superintegrability and associated polynomial solutions. Euclidean space and
the sphere in two dimensions; {\it J.Math.Phys.} {\bf 37}, 6439, (1996).
%-----------------------------------------------------------------------
\end{thebibliography}
\end{document}